# Disjoint LDPC Coding for Gaussian Broadcast Channels

Mahdi Ramezani and Masoud Ardakani

Department of Electrical and Computer Engineering, University of Alberta, Canada
Email: {ramezani,ardakani}@ece.ualberta.ca

*Abstract*—Low-density parity-check (LDPC) codes have been used for communication over a two-user Gaussian broadcast channel. It has been shown in the literature that the optimal decoding of such system requires joint decoding of both user messages at each user. Also, a joint code design procedure should be performed. We propose a method which uses a novel labeling strategy and is based on the idea behind the bit-interleaved coded modulation. This method does not require joint decoding and/or joint code optimization. Thus, it reduces the overall complexity of near-capacity coding in broadcast channels. For different rate pairs on the boundary of the capacity region, pairs of LDPC codes are designed to demonstrate the success of this technique.

## I. INTRODUCTION

The problem of simultaneous communication of a single source to multiple receivers, which is known as the broadcast channel, was first introduced by Cover in [1]. So far, the capacity region of certain classes of broadcast channels are known; however, the capacity region of a broadcast channel in general is still unknown.

Based on the achievable rate region given in [2], Berlin and Tuninetti in [3] studied the code design problem for a two-user fading Gaussian broadcast channel. They used low-density parity-check (LDPC) codes as the coding framework and studied the optimal decoding procedure. In fact, they showed that using superposition encoding and joint decoding, close-to-capacity LDPC codes can be found at low signal-to-noise ratios (SNRs). In their optimal scheme, since the user messages are superimposed, a joint factor graph is formed. Moreover, to obtain the maximum a posteriori (MAP) estimation of bits, the message updating rules impose mapper nodes. These mapper nodes connect the Tanner graphs associated with each user's code to act as interference cancelers. These mapper nodes not only increase the decoding complexity, they also require both users to have the codebook of each other and perform joint decoding. In addition, for near capacity performance, the codes should be jointly optimized. Most LDPC code design techniques are based on a search in the space of code parameters. The complexity of such search-based code design techniques increases significantly with the number of design parameters. A joint code optimization means that the number of design parameters is almost doubled.

In this paper, motivated by the bit-interleaved coded modulation (BICM) scheme [4], we propose a suboptimal scheme which employs a novel labeling method in order to remove the mapper nodes. Therefore, each user can use its own LDPC code. In other words, without the mapper nodes, there is no need for joint decoding; the users do not need to have the codes of each other and joint optimization of codes is not necessary. Our numerical results show that the proposed method performs close to the optimal solution of [3] (albeit with a lower decoding complexity.)

Since LDPC decoders usually use the log-likelihood ratio (LLR) values, we study the properties of LLRs based on the proposed method by a discussion on computing the probability density function (pdf) of LLRs.

In Section II, we briefly review the main results known for broadcast channels. We discuss using LDPC codes for a two-user Gaussian broadcast channel in Section III. Our method is proposed in Section IV and LDPC codes based on our method are designed in Section V. Section VI concludes the paper.

## II. BACKGROUND

A two-user broadcast channel consists of an input alphabet $\mathcal{X}$, two output alphabets $\mathcal{Y}$ and $\mathcal{Z}$, and a set of channel transition probabilities $p(y, z|x)$ where $(x, y, z) \in \mathcal{X} \times \mathcal{Y} \times \mathcal{Z}$. A $(2^{nR_y}, 2^{nR_z}, n)$ broadcast code consists of two equiprobable message sets $\mathcal{W}_y = \{1, 2, \ldots, M_y\}$ and $\mathcal{W}_z = \{1, 2, \ldots, M_z\}$ where $M_y = 2^{nR_y}$ and $M_z = 2^{nR_z}$, a codebook which has $M_y \times M_z$ codewords of length $n$ and symbols from the input alphabet $\mathcal{X}$, and two decoders which assign two message indices $\hat{w}_y(y^n) \in \mathcal{W}_y$ and $\hat{w}_z(z^n) \in \mathcal{W}_z$ to each received observation pair $(y^n, z^n)$ [1].

The goal is to send private messages to both receivers with a vanishing probability of error. The users can have a common message; however, in this work, we are only interested in the private messages. Two private messages are drawn independently from two message sets $\mathcal{W}_y$ and $\mathcal{W}_z$, and then the corresponding codeword is transmitted over the broadcast channel.

A pair of rates $(R_y, R_z)$ is said to be achievable if there exists a $(2^{nR_y}, 2^{nR_z}, n)$ broadcast code with vanishing average probabilities of error at both of the receivers, as $n \to \infty$ [1]. The capacity region of a broadcast channel is the convex closure of all the achievable rates and it depends only on the marginal densities, i.e., $p(y|x)$ and $p(z|x)$ [1].

The single-letter characterization of the capacity region of a general broadcast channel is still unknown. In special cases, however, the capacity region is known. Here, we confine our attention to degraded broadcast channels. If the broadcast channel transition probability can be factorized as $p(y,z|x) = p(z|y)p(y|x)$, then the broadcast channel is physically degraded which implies that $X$, $Y$, and $Z$ forms a Markov chain, i.e., $X \to Y \to Z$ [5]. In other words, user $Z$ receives a more degraded version of $X$ than user $Y$. We denote the convex hull by CH. Bergmans [6] proved that the capacity region of a degraded broadcast channel $X \to Y \to Z$ is the set of rates $(R_y, R_z)$ such that

$$\operatorname*{CH}_{p(v)p(x|v)} \left\{ R_y, R_z \geq 0 \;\middle|\; \begin{array}{l} R_z \leq I(V;Z) \\ R_y \leq I(X;Y|V) \end{array} \right\} \quad (1)$$

where $V$ is an auxiliary random variable whose support set $\mathcal{V}$ satisfies $|\mathcal{V}| \leq \min\{|\mathcal{X}|, |\mathcal{Y}|, |\mathcal{Z}|\}$. The idea is that the auxiliary random variable $V$ serves as a cloud center (cloud of codewords) distinguishable by both receivers [6]. There are in total $M_z$ clouds available and each cloud contains $M_y$ codewords. The "weaker" user, i.e., $Z$, can only see the clouds while the user $Y$ can also see codewords within a cloud. In fact, user $Y$ first strips off the message of user $Z$ (decodes the cloud) and then can see the individual codewords within a cloud [6]. This method is called *superposition coding*.

### A. Gaussian Broadcast Channels

The focus of this work is on the Gaussian broadcast channels which are defined as [3]

$$\begin{aligned} Y &= AX + N_y \\ Z &= BX + N_z \end{aligned} \quad (2)$$

where the additive white Gaussian noises are zero mean and have variance $N_0$, i.e., $N_y, N_z \sim \mathcal{N}(0, N_0)$, independent from the input $X$ which is power constrained by $\mathbb{E}(|X|^2) \leq P$. Also, $A$ and $B$ are two ergodic memoryless processes, known at the receivers. In general, the broadcast channel given in (2) is neither degraded nor more capable [2]. However, if the fading processes are constant (unfaded Gaussian) and $|A| > |B|$ then (2) will be degraded and the capacity region according to (1) is given by [3]

$$\bigcup_{\alpha \in [0,1]} \left\{ R_y, R_z \geq 0 \;\middle|\; \begin{array}{l} R_y \leq C(\alpha |A|^2 \gamma) \\ R_z \leq C(|B|^2 \gamma) - C(\alpha |B|^2 \gamma) \end{array} \right\}$$

where $C(x) = \frac{1}{2}\log_2(1+x)$ and $\gamma = \frac{P}{N_0}$. The boundary of this region is achieved by

$$X = \sqrt{\alpha P} U + \sqrt{\bar{\alpha} P} V \quad (3)$$

where $\mathbb{E}(|X|^2) = P$ and $\bar{\alpha} = 1 - \alpha$ for $\alpha \in [0,1]$. Also, $\alpha$ represents the fraction of power allocated for user $Y$, and $(U, V) \sim \mathcal{N}(\mathbf{0}, \mathbf{I}_2)$ is a pair of independent normal random variables [7].

Since the Gaussian input given in (3) cannot be used in practice, Berlin and Tuninetti in [3] consider the performance achievable by a binary linear codebook instead of the Gaussian codebook. This means that $(U, V)$ in (3) are now drawn uniformly from $\{-1, +1\} \times \{-1, +1\}$. In this case, the capacity region is given by [3]

$$\bigcup_{\alpha \in [0,1]} \left\{ R_y, R_z \geq 0 \;\middle|\; \begin{array}{l} R_y \leq J(\alpha |A|^2 \gamma) \\ R_z \leq J(|B|^2 \gamma) - J(\alpha |B|^2 \gamma) \end{array} \right\}$$

where $J(t) = 1 - \mathbb{E}_M \log_2(1 + e^{-M})$ and $M \sim \mathcal{N}(t/2, t)$.

We use LDPC codes in this work as the binary code to communicate over the Gaussian broadcast channel.

### B. LDPC Codes

In this paper, following the notation of [8], an ensemble of LDPC codes is defined by a pair of distributions $(\lambda, \rho)$ in the polynomial form, i.e., $\lambda(x) = \sum_{i \geq 2} \lambda_i x^{i-1}$, and $\rho(x) = \sum_{i \geq 2} \rho_i x^{i-1}$. Throughout this work, we will use appropriate subscripts to distinguish the codes of different users. Transmission of LDPC codes takes place on a memoryless binary-input symmetric-output (BISO) channel. Under the sum-product decoding and the all-one codeword[1] transmission assumption [8], a BISO channel is completely characterized by the pdf of its LLR messages, denoted by $\mathsf{a}_{\mathrm{ch}}(x)$ where for all $x$, we have $\mathsf{a}_{\mathrm{ch}}(-x) = \mathsf{a}_{\mathrm{ch}}(x)e^{-x}$.

By code design for a fixed channel, we mean optimizing the polynomials $\lambda(x)$ and $\rho(x)$ in order to obtain an LDPC code with the highest code rate which converges to zero error-rate on the given channel. Similar to [8], we fix $\rho(x)$ so that the optimization problem can be formulated as a linear programming.

## III. LDPC Coding for Gaussian Broadcast Channels

For LDPC coding on the broadcast channel, as suggested in [3], one can pick two LDPC codes from the ensembles $(\lambda_y, \rho_y)$ and $(\lambda_z, \rho_z)$. Denoting codewords of length $n$ by boldface letters, the superimposed transmitted vector $\mathbf{x}$ can be written as

$$\mathbf{x} = \sqrt{\alpha P}\mathbf{x}_y + \sqrt{\bar{\alpha} P}\mathbf{x}_z \quad (4)$$

where $\mathbf{x}_y, \mathbf{x}_z \in \{-1, +1\}^n$ are the binary codewords of the users $Y$ and $Z$, respectively. Also, by transmission of $\mathbf{x}$, we observe two vectors $\mathbf{y}$ and $\mathbf{z}$.

The factor graph associated with the MAP estimate of $x_{i,y}$, the $i$th bit of the binary vector $\mathbf{x}_y$, is given by [3] where the function node connecting the two Tanner graphs is called the mapper node. The mapper node increases the complexity in the following ways:

- The mapper node forces both users to have the code of each other in order to jointly decode the codewords.
- The decoding complexity increases considerably. This is because at each iteration of the message passing decoder, the messages from one Tanner graph should be passed to the other graph to enhance the reliability of decisions.
- The code design stage becomes more cumbersome than in the single-user case as the codes must be optimized jointly.

---
[1]We use the conventional mapping $0 \mapsto 1$ and $1 \mapsto -1$.

## IV. A Disjoint LDPC Coding Scheme

In this section, we propose a method to use LDPC codes over a two-user Gaussian broadcast channel based on a novel adaptive labeling method and interleaving the bits. The main goal is to allow for disjoint decoding of messages and the idea is motivated by BICM. Thus, a brief discussion on BICM which leads to our main idea is presented.

### A. Bit-Interleaved Coded Modulation

BICM is a bandwidth-efficient coding method [4] which is based on the concatenation of a binary code, a bit-interleaver and a high-order modulation [9]. The coded bits in BICM are interleaved and then every $d = \log_2 D$ bits are grouped together and sent over the channel using a $D$-ary constellation. At the receiver, after computing the LLR values of the coded bits and de-interleaving, a binary decoder is used as if the LLR values were the observations at a binary phase shift keying channel output [9].

Now, consider two labeling methods: Gray labeling and binary set partitioning [10] (Ungerboeck) labeling. In Gray labeling, the label of each point of the constellation differs from its neighbors only in one bit. Ungerboeck labeling partitions the constellation such that each bit has a different level of protection and the received symbols have to be decoded sequentially. It has been shown [4] that if we use Gray labeling for the $D$-ary constellation, then the capacity of BICM is extremely close to the capacity of the optimal receiver. Using binary set partitioning (Ungerboek labeling) and BICM, the capacity is far from the optimal decoding since the BICM receiver considers that the bits are independent while the optimal receiver exploits the dependency between the successive bits.

An important result of [4] is that using a bit interleaver and Gray labeling, a binary decoder can be used to get a performance which is almost the same as the optimal receiver. Another way of understanding this result is that the dependency among the label bits in a Gray-labeled constellation is minor, allowing the decoder not to exploit the dependency and still provide near-optimal performance.

### B. The Proposed Method

Let us have a look at the superimposed codeword given in (4). The transmitted symbol $X$ is selected from the set $\mathcal{X} = \{\pm\sqrt{\alpha P} \pm \sqrt{\bar{\alpha} P}\}$ that can be viewed as a mapping which maps two independent bits to a point in a 4-PAM-like constellation shown in Table I. This mapping uses a binary labeling method and it should be emphasized that depending on the value of $\alpha$, we can have different configurations. In the first column of Table I, the most significant bit position represents $X_z$ and the other bit represents $X_y$.

Now, consider two sequences of LDPC coded bits, each of which is intended for one of the users. We utilize the fact that LDPC codes are self-interleaved and apply Gray labeling for the 4-PAM-like constellation. Table I shows bit configurations for this scheme. When $\alpha \geq \frac{1}{2}$, the transmitted codeword has to be

$$X = \sqrt{\alpha P} X_y + \sqrt{\bar{\alpha} P} X_y X_z \quad (5)$$

in order to have Gray labeling. Since $X_y$ and $X_z$ are independent, zero-mean and unit-variance binary random variables, we obtain $\mathbb{E}(|X|^2) = \alpha P \mathbb{E}(|X_y|^2) + \bar{\alpha} P \mathbb{E}(|X_z|^2) \mathbb{E}(|X_y|^2) = P$ which shows that the power constraint is satisfied.

To maintain Gray labeling for $\alpha \leq \frac{1}{2}$, as it is shown in Table I, the position of bits has to be interchanged. The corresponding transmitted codeword is

$$X = \sqrt{\alpha P} X_y X_z + \sqrt{\bar{\alpha} P} X_z \quad (6)$$

which satisfies the power constraint as $\mathbb{E}(|X|^2) = \alpha P \mathbb{E}(|X_y|^2) \mathbb{E}(|X_z|^2) + \bar{\alpha} P \mathbb{E}(|X_z|^2) = P$.

*Remark 1:* It is noteworthy that the two parts of (5) and (6) are not independent since they share a factor in common. Therefore, our method does not exactly match superposition coding. Also, in superposition coding, the message intended for the weaker user identifies the cloud centers, no matter how much power we allocate to the weaker user. From (5), i.e., for $\alpha \geq \frac{1}{2}$, the cloud center is not identified by the message of user $Z$, i.e., $X_y X_z$. Therefore, the region obtained based on the proposed method may not be convex. This is because we adaptively force the labeling to be Gray, leading to a mismatch between the proposed method and superposition coding. However, we will see in Section V that the region based on our method covers most of the region given in [3]. ▲

As mentioned, by using Gray labeling we have reduced the dependency among the label bits. Interleaving removes the dependency altogether to validate our decoding approach. Moreover, interleaving does not incur a significant loss since the dependency among the bits was indeed minor due to the labeling scheme. An optimal decoder should still use the existing dependency, but the performance gain will be minor. In binary labeling, the dependency must be used (that is what the mapper node does) because it is too strong.

To analyze the proposed method, let us determine the capacity region using our method. For $\alpha \geq \frac{1}{2}$, we have

$$R_z \leq I(V; Z)$$
$$= \sum_{x_z \in \{\pm 1\}} \int p(x_z) p(z|x_z) \log_2 \frac{p(z|x_z)}{p(z)} dz$$
$$= 1 - \frac{1}{2} \int p(z|X_z = +1) \log_2 \left(1 + \frac{p(z|X_z = -1)}{p(z|X_z = +1)}\right) dz -$$
$$\frac{1}{2} \int p(z|X_z = -1) \log_2 \left(1 + \frac{p(z|X_z = +1)}{p(z|X_z = -1)}\right) dz$$

and

$$R_y \leq I(X; Y|V)$$
$$= H(Y|X_z) - H(Y|X, X_z)$$
$$= H(Y|X_z) - H(AX + N_y|X, X_z)$$
$$= \frac{1}{2}\left[H(Y|X_z = +1) + H(Y|X_z = -1)\right] - \frac{1}{2}\log_2(2\pi e N_0)$$

TABLE I: Comparison of binary and Gray labeling methods where the symbol $P$ is removed for simplicity. Note that for binary labeling, we always have $X = \sqrt{\alpha P} X_y + \sqrt{\bar{\alpha} P} X_z$.

| Binary labeling | Gray labeling | $\alpha \in [0,1]$ |
|---|---|---|
| 11, 01, 10, 00 at $-\sqrt{\alpha}-\sqrt{\bar\alpha}$, $\sqrt{\bar\alpha}-\sqrt{\alpha}$, $\sqrt{\alpha}-\sqrt{\bar\alpha}$, $\sqrt{\alpha}+\sqrt{\bar\alpha}$ ($X_z, X_y$ at 00) | $X = \sqrt{\alpha P} X_y + \sqrt{\bar{\alpha} P} X_z X_y$; 01, 11, 10, 00 at $-\sqrt{\alpha}-\sqrt{\bar\alpha}$, $\sqrt{\bar\alpha}-\sqrt{\alpha}$, $\sqrt{\alpha}-\sqrt{\bar\alpha}$, $\sqrt{\alpha}+\sqrt{\bar\alpha}$ ($X_z, X_y$ at 00) | $\alpha \geq \tfrac{1}{2}$ |
| 11, 10, 01, 00 at $-\sqrt{\alpha}-\sqrt{\bar\alpha}$, $\sqrt{\bar\alpha}-\sqrt{\alpha}$, $\sqrt{\alpha}-\sqrt{\bar\alpha}$, $\sqrt{\alpha}+\sqrt{\bar\alpha}$ ($X_z, X_y$ at 00) | $X = \sqrt{\alpha P} X_y X_z + \sqrt{\bar{\alpha} P} X_z$; 01, 11, 10, 00 at $-\sqrt{\alpha}-\sqrt{\bar\alpha}$, $\sqrt{\bar\alpha}-\sqrt{\alpha}$, $\sqrt{\alpha}-\sqrt{\bar\alpha}$, $\sqrt{\alpha}+\sqrt{\bar\alpha}$ ($X_y, X_z$ at 00) | $\alpha \leq \tfrac{1}{2}$ |

where $p(z|x_z = +1)$ is a mixture of two Gaussian pdfs. In a similar manner, we can derive formulas for $\alpha \leq \tfrac{1}{2}$. Finally, the region based on the proposed method is given by (1).

### C. LLR Computation and Its Properties

Usually LDPC decoders use LLR values. In this section we study some of the properties of LLRs for our proposed solution. In particular, we see that the LLR pdf does not satisfy the symmetry required for the all-one codeword assumption. We first describe how to obtain the LLRs for each of the users. For user $Y$ and $\alpha \geq \tfrac{1}{2}$, according to Table I, the LLR message received from the channel $p(y|x_y)$ is

$$m_y = \log_2 \frac{p(y|X_y = +1)}{p(y|X_y = -1)}$$
$$= \log_2 \frac{\sum_{x_z} p(x_z) p(y|X_y = +1, x_z)}{\sum_{x_z} p(x_z) p(y|X_y = -1, x_z)}$$
$$= \log_2 \frac{\Omega_A(+1,+1) + \Omega_A(+1,-1)}{\Omega_A(-1,-1) + \Omega_A(-1,+1)}$$

where

$$\Omega_A(p,q) = \frac{1}{\sqrt{2\pi N_0}} \exp\left\{\frac{-1}{2N_0}\big(y - A\sqrt{P}(p\sqrt{\alpha}+q\sqrt{\bar\alpha})\big)^2\right\}$$

and $p, q \in \{\pm 1\}$. For $\alpha \leq \tfrac{1}{2}$, we obtain

$$m_y = \log_2 \frac{p(y|X_y = +1)}{p(y|X_y = -1)}$$
$$= \log_2 \frac{\Omega_A(+1,+1) + \Omega_A(-1,-1)}{\Omega_A(-1,+1) + \Omega_A(+1,-1)}.$$

Similarly, for user $Z$, we get

$$m_z = \log_2 \frac{\Omega_B(+1,+1) + \Omega_B(-1,-1)}{\Omega_B(+1,-1) + \Omega_B(-1,+1)}$$

and

$$m_z = \log_2 \frac{\Omega_B(+1,+1) + \Omega_B(-1,+1)}{\Omega_B(-1,-1) + \Omega_B(+1,-1)},$$

for $\alpha \geq \tfrac{1}{2}$ and $\alpha \leq \tfrac{1}{2}$, respectively.

*Lemma 1:* The pdfs $p(y|x_y)$ and $p(z|x_z)$ are not symmetric.

*Proof:* We have

$$p(y|x_y) = \sum_{x_z} p(x_z) p(y|x_y, x_z)$$
$$= \frac{1}{2}\big[p(y|x_y, X_z = +1) + p(y|x_y, X_z = -1)\big].$$

For $\alpha \geq \tfrac{1}{2}$, we get $p(y|X_y = -1) = \tfrac{1}{2}\big[\Omega_A(-1,-1) + \Omega_A(-1,+1)\big] = p(-y|X_y = +1)$, but this does not hold for $\alpha \leq \tfrac{1}{2}$ since $p(y|X_y = -1) = \tfrac{1}{2}\big[\Omega_A(-1,+1) + \Omega_A(+1,-1)\big] \neq p(-y|X_y = +1)$. This result can be extended to $p(z|x_z)$. ∎

Let us denote the pdf of $m_y$ by $\mathrm{a}_{\mathrm{ch},y}(m)$. Since the channel $p(y|x_y)$ is not symmetric, the all-one codeword assumption for the purpose of density evolution [11] is not valid. However, according to [12], a symmetrized LLR pdf can be obtained using

$$\mathrm{a}_{\mathrm{ch},y}(m) = \frac{1}{2}\big[\mathrm{a}_{\mathrm{ch},y}(m|X_y = +1) + \mathrm{a}_{\mathrm{ch},y}(-m|X_y = -1)\big].$$

Note that if the channel was symmetric, the above equation would lead to $\mathrm{a}_{\mathrm{ch},y}(m) = \mathrm{a}_{\mathrm{ch},y}(m|X_y = +1)$, resulting in the all-one codeword assumption. Similarly, we denote the LLR pdf of user $Z$ by $\mathrm{a}_{\mathrm{ch},z}(m)$.

## V. SIMULATION RESULTS AND CODE DESIGN

In this section, we compare the regions resulting from a Gaussian input, binary labeling method from [3] and our proposed method. Then, we design LDPC codes based on our labeling method.

Consider a two-user Gaussian broadcast channel given in (2) where $|A|^2 \gamma = 5.059$ dB, and $|B|^2 \gamma = 3.871$ dB. We choose these values to have a fair comparison with the region given in [3]. In Fig. 1, we compare the capacity region when the optimal Gaussian input is used, the optimal region given in [3], and the region based on our method in Section IV. It can be seen that most of the optimal region is covered by our proposed method. As we discussed in Remark 1, the region based on our method is not convex. Since the proposed method does not require joint decoding, for each user a separate LDPC code can be optimized using the conventional techniques in the literature. In our numerical optimizations, we use the sum-product discrete density evolution established by Chung in [13], where the maximum LLR is set to 25 and a 9-bit quantizer is used. We allow codes with maximum variable node degree of 50 converging in at most 800 iterations to a target bit error rate of $10^{-6}$. It is noteworthy that the LLR pdf for each user is obtainable using the discussion in Section IV.

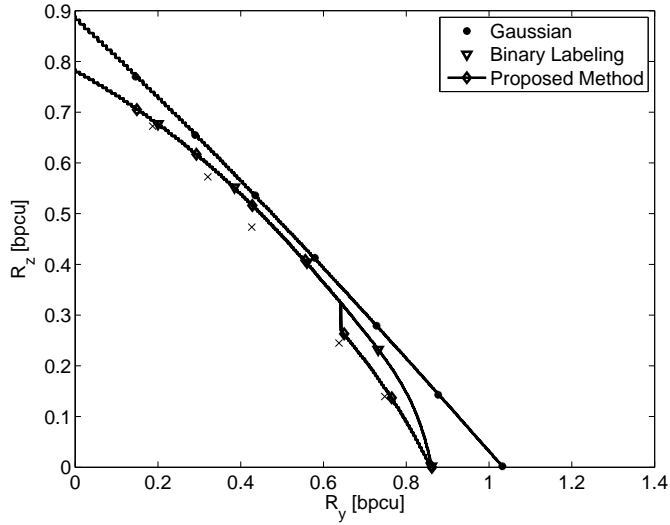

Fig. 1: Comparison of the capacity region of a two-user Gaussian broadcast channel with different inputs. The cross points show the achieved rates by the proposed method given in Table II and Table III.

TABLE II: Optimized degree distributions for user $Y$.

| $\alpha$ | 0.1 | 0.2 | 0.3 | 0.8 |
|---|---|---|---|---|
| $\lambda_2, \lambda_3$ | 0.276, 0.215 | 0.263, 0.233 | 0.190, 0.201 | 0.162, 0.194 |
| $\lambda_5, \lambda_6$ |  | 0.000, 0.009 | 0.007, 0.029 | 0.007, 0.035 |
| $\lambda_7, \lambda_8$ | 0.012, 0.210 | 0.144, 0.071 | 0.055, 0.045 | 0.065, 0.036 |
| $\lambda_9, \lambda_{10}$ | 0.012, 0.020 | 0.014, 0.007 | 0.042, 0.044 | 0.030, 0.033 |
| $\lambda_{11}, \lambda_{12}$ |  | 0.005, 0.029 | 0.036, 0.021 | 0.037, 0.031 |
| $\lambda_{13}, \lambda_{14}$ |  |  | 0.012, 0.007 | 0.020, 0.012 |
| $\lambda_{15}, \lambda_{16}$ |  |  | 0.005, 0.036 | 0.007, 0.033 |
| $\lambda_{19}$ |  | 0.006 |  |  |
| $\lambda_{20}, \lambda_{21}$ |  | 0.007, 0.009 |  |  |
| $\lambda_{22}, \lambda_{23}$ |  | 0.011, 0.025 |  |  |
| $\lambda_{24}, \lambda_{25}$ |  | 0.018, 0.022 |  |  |
| $\lambda_{26}, \lambda_{27}$ |  | 0.023, 0.022 |  |  |
| $\lambda_{28}, \lambda_{29}$ |  | 0.018, 0.014 |  |  |
| $\lambda_{30}, \lambda_{31}$ |  | 0.011, 0.008 |  |  |
| $\lambda_{32}, \lambda_{33}$ |  | 0.006, 0.036 |  |  |
| $\lambda_{48}, \lambda_{49}$ |  |  | 0.000, 0.009 | 0.025, 0.010 |
| $\lambda_{50}$ | 0.249 |  | 0.261 | 0.263 |
| $\rho_5, \rho_6$ | 1, 0 | 0.173, 0.827 |  |  |
| $\rho_8, \rho_9$ |  |  | 0.471, 0.529 |  |
| $\rho_{14}, \rho_{15}$ |  |  |  | 0.483, 0.517 |
| Rate | 0.187 | 0.320 | 0.426 | 0.637 |

TABLE III: Optimized degree distributions for user $Z$.

| $\alpha$ | 0.1 | 0.2 | 0.3 | 0.8 |
|---|---|---|---|---|
| $\lambda_2, \lambda_3$ | 0.148, 0.197 | 0.169, 0.200 | 0.238, 0.237 | 0.292, 0.235 |
| $\lambda_5, \lambda_6$ | 0.007, 0.027 | 0.007, 0.024 | 0.000, 0.011 | 0.000, 0.009 |
| $\lambda_7, \lambda_8$ | 0.067, 0.047 | 0.057, 0.050 | 0.198, 0.025 | 0.096, 0.101 |
| $\lambda_9, \lambda_{10}$ | 0.035, 0.032 | 0.043, 0.040 | 0.008, 0.005 | 0.018, 0.009 |
| $\lambda_{11}, \lambda_{12}$ | 0.030, 0.026 | 0.033, 0.023 |  | 0.006, 0.005 |
| $\lambda_{13}, \lambda_{14}$ | 0.019, 0.013 | 0.014, 0.009 |  |  |
| $\lambda_{15}, \lambda_{16}$ | 0.009, 0.006 | 0.006, 0.038 | 0.045, 0.007 |  |
| $\lambda_{17}, \lambda_{18}$ | 0.005, 0.049 |  | 0.007, 0.009 |  |
| $\lambda_{19}, \lambda_{20}$ |  |  | 0.014, 0.023 | 0.060, 0.008 |
| $\lambda_{21}, \lambda_{22}$ |  |  | 0.038, 0.049 | 0.005, 0.006 |
| $\lambda_{23}, \lambda_{24}$ |  |  | 0.037, 0.022 | 0.006, 0.007 |
| $\lambda_{25}, \lambda_{26}$ |  |  | 0.013, 0.008 | 0.008, 0.009 |
| $\lambda_{27}, \lambda_{28}$ |  |  | 0.006, 0.000 | 0.011, 0.012 |
| $\lambda_{29}, \lambda_{30}$ |  |  |  | 0.012, 0.013 |
| $\lambda_{31}, \lambda_{32}$ |  |  |  | 0.012, 0.012 |
| $\lambda_{33}, \lambda_{34}$ |  |  |  | 0.011, 0.010 |
| $\lambda_{35}, \lambda_{36}$ |  |  |  | 0.008, 0.007 |
| $\lambda_{37}, \lambda_{38}$ |  |  |  | 0.006, 0.006 |
| $\lambda_{47}, \lambda_{48}$ | 0.005, 0.008 | 0.018, 0.006 |  |  |
| $\lambda_{49}, \lambda_{50}$ | 0.015, 0.255 | 0.011, 0.252 |  |  |
| $\rho_4, \rho_5$ |  |  |  | 0, 1 |
| $\rho_7, \rho_8$ |  | 0.273, 0.727 |  |  |
| $\rho_{12}$ |  | 1 |  |  |
| $\rho_{16}, \rho_{17}$ | 0.485, 0.515 |  |  |  |
| Rate | 0.672 | 0.572 | 0.473 | 0.244 |

In Fig. 1, the achievable rates are shown by the cross points. Also, due to the limit of space, some of the optimized degree distributions for user $Y$ and $Z$ are reported in Table II and Table III, respectively.

## VI. CONCLUSION

In this paper, a low complexity method for communicating over a two-user Gaussian broadcast channel based on LDPC codes was presented. Compared to the previous works, we showed that in our method, each user can use a single LDPC code and the need for joint decoding at the receivers is eliminated.